\documentclass[12pt,preprint]{aastex}
\usepackage{longtable}
\usepackage{graphicx}
\usepackage{subfigure}
\usepackage{longtable}
\usepackage{indentfirst}
\usepackage{amsmath}

\shorttitle{Modelling KIC 8263801}
\shortauthors{Tang et al.}

\begin{document}

\title{Asteroseismology of KIC 8263801:\\
Is it a member of NGC 6866 and a red clump star?}

\author{Yanke Tang\altaffilmark{1,2,7}{$\star$},
Sarbani Basu\altaffilmark{2}{$\star$},
Guy R.~Davies\altaffilmark{3},
Earl P.~Bellinger\altaffilmark{4,2,5,6},
Ning Gai\altaffilmark{1,2,7}}

\affil{\altaffilmark{1} College of Physics and Electronic information, Dezhou University, Dezhou 253023, China \\
\altaffilmark{2} Department of Astronomy, Yale University, P.O. Box 208101, New Haven, CT 06520-8101, USA \\
\altaffilmark{3} School of Physics and Astronomy, University of Birmingham, Birmingham, B15 2TT, United Kingdom \\
\altaffilmark{4} Max-Planck-Institut f{\"u}r Sonnensystemforschung, Justus-von-Liebig-Weg 3, 37077 G{\"o}ttingen, Germany \\
\altaffilmark{5} Stellar Astrophysics Centre, Department of Physics and Astronomy, Aarhus University, Ny Munkegade 120, DK-8000 Aarhus C, Denmark \\
\altaffilmark{6} Institut f{\"u}r Informatik, Georg-August-Universit{\"a}t G{\"o}ttingen, Goldschmidtstrasse 7, 37077 G{\"o}ttingen, Germany \\
\altaffilmark{7} Shandong Provincial Key Laboratory of Biophysics, Dezhou University, Dezhou 253023, China \\
$^\star$ Corresponding authors, email: tyk@dzu.edu.cn; sarbani.basu@yale.edu; ning.gai@hotmail.com}

\begin{abstract}
We present an asteroseismic analysis of the {\it Kepler} light curve of KIC
8263801, a red-giant star in the open cluster NGC 6866 that has previously been
reported to be a helium-burning red-clump star.  We extracted the frequencies
of the radial and quadrupole modes from its frequency power spectrum and
determined its properties using a grid of evolutionary models constructed with
MESA. The oscillation frequencies were calculated using the GYRE code and the
surface term was corrected using the \cite{Ball14} prescription.  We find that
the star has a mass of $M/M_{\odot} = 1.793\pm 0.072$, age $t=1.48\pm 0.21$ Gyr and
radius $R/R_{\odot} = 10.53\pm 0.28$.  By analyzing the internal structure of the best-fitting
model, we infer the evolutionary status of the star KIC~8263801 as being on the
ascending part of the red giant branch, and not on the red clump.  This result
is verified using a purely asteroseismic diagnostic, the
$\epsilon_{c}-\Delta\nu_{c}$ diagram which can distinguish red giant branch
stars from red clump stars.  Finally, by comparing its age with NGC 6866
($t=0.65 \pm 0.1$ Gyr) we conclude that KIC 8263801 is not a member of this
open cluster.
\end{abstract}

\keywords{stars: evolution -- stars: red giant -- stars: oscillations}

\section{Introduction}
Since the NASA {\it Kepler} spacecraft was successfully launched in March 2009,
stellar pulsations of over 10,000 red giants have been
observed \citep{Borucki09, Borucki10, de ridder09,
bedding10, koch10, gilliland10, Balona13a, Balona13b, Abedigamba16}.
The red giant (RG) star KIC 8263801 is one of these.  Located in the field of
NGC 6866---the youngest of the four clusters observed by {\it Kepler}---the pulsations of KIC 8263801 have been observed with a signal-to-noise ratio that is good enough  to constrain its fundamental parameters.
\citet{Balona13b} identified 704 stars in this cluster, of which 23 are RG stars showing solar-like oscillations.
\citet{Abedigamba16} used
median gravity-mode period spacings ($\Delta P$) to search for red clump (RC)
stars among the RG stars showing solar-like oscillations.  Based on
its $\Delta P = 173.7 \pm 6.4$ s, \citet{Abedigamba16} determined that KIC 8263801 is a secondary red clump (SRC) star, which is massive enough to have ignited helium
burning in a non-degenerate core.
However, the signal-to-noise of KIC 8263801 is not enough to derive an accurate
$\Delta P$.  In this paper, we perform an asteroseismic analysis of the
mode frequencies in order to determine the evolutionary state of
KIC 8263801.

One of the major problems in the study of open clusters is to determine which
stars are members of the clusters and which stars are not.
Although, KIC 8263801 has been classified as
non-member of NGC 6866 based on photometric distance membership determination
and proper motions \citep{Balona13b}, \cite{Balona13b} and
\cite{Abedigamba16} pointed out that they can not be quite sure of the
membership with only a single method.  Abedigamba (2016) stated that NGC6866 is roughly located in the direction of solar apex, which means that all stars (members and non-members) have similar proper motion. one cannot use proper motion to discriminate between members and non-members. No radial velocity measurements for stars located in the field of NGC6866 are available to help in discriminating between members and non-members. The only method left in discriminating between members and non-members in the field of NGC6866 is the photometric distance method. However, with only a single method one cannot be quite sure of the membership. Balona et al. (2013) also attempted to identify cluster members using their proper motions but found very poor discrimination between members and non-members.
Hence, another focus of this paper is thus to
revisit the issue of the cluster non-membership of KIC 8263801 by means of
comparing the age of the star determined through asteroseismic modelling and the
age of the cluster.

Asteroseismology is an efficient tool for
studying the internal structures of stars through their global oscillations
\citep{tang08a, tang08b, tang11, Chaplin11,Chaplin14, Chaplin14b, Pinsonneault14,Silva15,Silva17,Bellinger16,Bellinger17}.
Basic stellar properties such as mass, age can be derived this way
\citep[e.g.,][]{basu10a, basu10b,Yildiz17}. For the stars with solar-like oscillations, when the signal-to-noise ratios in the seismic data are insufficient to allow robust extraction of individual oscillation frequencies, it is still possible to extract estimates of  the large frequency separation ($\Delta\nu$) and the frequency of
maximum oscillation power ($\nu_{\max}$).  The large separation $\Delta\nu$ is
the separation between oscillation modes with the same angular degree and
consecutive radial orders \citep{Tassoul80}:
\begin{equation} \label{eq:lsp}
    \Delta\nu_{l}(n)=\nu_{n,l}-\nu_{n-1,l}
\end{equation}
which scales to a very good approximation with the square root of the stellar
mean density \citep{Kjeldsen95,stello11}:
\begin{equation} \label{eq:lsp2}
    \frac{\Delta \nu}{\Delta\nu_{\odot}}\simeq\sqrt{\frac{M/M_{\odot}}{(R/R_{\odot})^{3}}}.
\end{equation}
The other, $\nu_{\max}$, is related with the cut-off frequency for acoustic
waves in an isothermal atmosphere, which scales with surface gravity $g$ and
effective temperature $T_{\rm{eff}}$ \citep{brown91, Kjeldsen95} as:
\begin{equation} \label{eq:ac}
    \frac{\nu_{\max}}{\nu_{max\odot}}
    \simeq
    \left(\frac{g}{g_{\odot}}\right)
    \left(\frac{T_{\rm{eff}}}{T_{\rm{eff},\odot}}\right)^{-1/2}
\end{equation}
with the solar values $\Delta\nu_{\odot}=135$ $\mu$Hz and $\nu_{\max\odot} =
3090$ $\mu$Hz \citep{huber09, huber11, Chaplin14} providing the absolute
calibration in this study.  These scaling relations have been applied with
success to main-sequence and sub-giant stars as well as to stars on the
red giant branch (RGB) and helium
burning (HeB) stars. According to the above scaling relations, asteroseismology can help constrain the global parameters of a star with reasonable precision. For more evolved stars, the scaling relation has reduced accuracy \citep{Gualme16,Guggenberger16,Guggenberger17}.

The classical Hertzsprung-Russell diagram (HRD) shows that red giant branch (stars burning hydrogen in a shell around an
inert helium core) and red clump (stars with Helium-core and hydrogen-shell burning) occupy
overlapping parameter spaces \citep[e.g.][]{Elsworth17}. Although these stars have different internal conditions, which we can observe indirectly, they have similar surface characteristics---such as effective temperature,
surface gravity and total luminosity, which can be observed directly.
Hence, from classical observations such as these, it is often not possible to
distinguish between RGB stars of about 10-12 $R_{\odot}$ and HeB stars of a
similar radius. The distinction is easier in clusters, but even in clusters
there is room for ambiguity. This is where asteroseismology is useful.
Space missions such as {\it Kepler} have made red-giant asteroseismology
possible, with {\it Kepler} providing long time series data that
allow the frequency resolution needed to determine frequencies
of individual oscillation modes of red giants \citep{bedding11,beck11,kallinger12,Handberg17}.
Non-radial modes in red giants exhibit mixed character; they behave like
gravity modes (g modes) in the interior but like acoustic modes (p modes) in
the outer layers. The asteroseismic properties of RGB and HeB stars
are quite different \citep{bedding11,kallinger12,Elsworth17} and
hence are useful in determining the evolutionary state of red giants. This what
we do in this paper. We extract the mode frequencies of the
radial and quadrupole modes KIC 8263801, construct models that match the frequencies
and use those to identify the evolutionary state
of the star to determine whether or not it is a member of NGC~6866.
Additionally, we also use a purely asteroseismic diagnostic to confirm its
evolutionary state. We do not use the dipole modes since the frequencies and
frequency spacings of the most g-type modes are determined by the details
of the profile of the buoyancy frequency which depends on uncertain model
parameters such as the exact treatment of overshoot etc. Radial ($l=0$) and quadrupole ($l=2$)
mode frequency are more robust in this respect.

In this paper, we extract the mode
frequencies of the radial and quadrupole modes by `peak bagging' the frequency
power spectrum observed by {\it Kepler} in \S~\ref{sec:obs}.  We use these individual mode
frequencies to constrain models of KIC 8263801, identify the evolutionary state
of the star, and confirm that it is not a member of the open cluster NGC 6866 in  \S~\ref{sec:model} and  \S~\ref{sec:results}.
Conclusions are presented in \S~\ref{sec:conc}.
\section{Observational data}
\label{sec:obs}

\subsection{Asteroseismic fitting}
\label{sec:asterob}

We extract the mode frequencies of the radial and quadrupole modes
using the KASOC unweighted power spectrum (Handberg \& Lund 2014) containing the Kepler
data spanning Quarters 1-3, 6-11, and 14 shown in Figure 1.
The data result in a frequency resolution of $\approx$ 9 nHz.
First guesses to the frequencies of the radial and quadrupole modes
were obtained using a fit of the asymptotic expression for radial
modes to the power spectrum (Davies \& Miglio 2016).
Peakbagging for these modes was does using the procedure, and the code, of
Davies et al. (2016). As is described in Davies et al. (2016), an unsupervised machine-learning Bayesian scheme is used to obtain
a set of probabilities to assess whether or not a mode had
been detected in the data. We use the ratio of the
probability of a detection over the probability of no detection (the
Bayes factor). If the natural log of the Bayes factor is high, we accept the mode. Only 5 radial orders passed the quality test.
The fitted frequencies are given in Table 1. We derive the values of large
separation $\Delta \nu =5.35\pm0.01\mu Hz$ and $\nu_{max}=56.2\pm0.4\mu Hz$ shown in Table 2, which are consistent
with the results obtained by Abedigamba (2016).

\subsection{Classical observations}
\label{sec:nonobs}

The effective temperature $T_{\rm{eff}}=4766$ K, metallicity
$[\text{Fe}/\text{H}] = 0.016$ dex, and surface gravity $\log g = 2.487$
(dex) of KIC 8263801 were obtained from \cite{brown11}. The stated
uncertainties of these quantities were 200 K in $T_{\rm{eff}}$ and 0.4 dex in
$\log g$. Although an uncertainty for the metallicity was not reported, the uncertainties were
expected to be high.
\cite{huber14} subsequently presented
revised stellar properties for 196\,468 {\it Kepler} targets, including this one.
They give $T_{\rm{eff}}=4974 \pm 161 K$, metallicity
$[\text{Fe}/\text{H}]= 0.016 \pm 0.30$ (dex), and $\log g= 2.662 \pm 0.03$
for this star.  We use these observational constraints as the error boxes as
shown in Figure~\ref{fig:track}, where maximum (black) uncertainty corresponds
to the observational value from the {\it Kepler} website and minimum (green)
uncertainty corresponds to the observational value given by \citet{huber14}.
Due to the location of KIC~8263801 in the field of NGC 6866, we also consider
that it may have a similar metallicity to this cluster.  \citet{Bostanci15}
and \citet{Balona13b} give the metallicity of the cluster as being about solar
value, whereas \citet{Loktin94} derived $[\text{Fe}/\text{H}] = 0.10$ (dex) via
photometry from \citet{Hoag61}.  All non-asteroseismic observational
constraints are listed in Table \ref{tab:classical-obs}.

\section{Modelling KIC 8263801}
\label{sec:model}

\subsection{Constructing models}
Having values of $\nu_{\max}$, $\Delta \nu$ and $T_{\rm{eff}}$, the scaling
relations (Equations~\ref{eq:lsp2} and~\ref{eq:ac}) can be used to calculate
the mass, luminosity, and radius of the star.
Using these scaling relations, \citet{Abedigamba16} determined that KIC 8263801
has a mass of $M/M_{\odot}=1.86$ and luminosity $\log (L/L_{\odot}) = 1.7555$.
However, more robust estimates of these quantities can be obtained by means of
a grid-based analysis, which constrains stellar parameters by searching among a
grid of evolutionary models to get a best fit for observed values of
$\nu_{\max}$, $\Delta \nu$, $T_{\rm{eff}}$ and metallicity
\citep[e.g.,][]{basu10a,basu11,gai11,Chaplin14,Chaplin14b}.
Since we have extracted the individual
frequencies of the star in addition to the global asteroseismic parameters, we
can use grid-based modelling to obtain better results by selecting models that
fit the observed frequencies and not just the
global asteroseismic parameters $\nu_{\max}$ and $\Delta \nu$.

We use the MESA stellar evolution
code \citep[\emph{Modules for Experiments in Stellar Astrophysics},][version
8845]{Paxton11,Paxton13} to construct models. We employ the default MESA input options unless
otherwise stated.  We evolve each pre-main sequence model until its nuclear
luminosity first reaches 99.9\% of the total luminosity, which we define as
being zero-age main sequence (ZAMS).  We use the Eddington T-$\tau$ atmosphere
from ZAMS onwards.  Each track is then evolved from ZAMS to asymptotic giant
branch (AGB) as shown in Figure~\ref{fig:track}.  We treat the convection
zone by the standard mixing-length theory (MLT) of \citet{Cox68} with the
mixing-length parameter $\alpha_{\text{MLT}}=2.0$ \citep{Wu16}.

Referring to the mass value derived by \citet{Abedigamba16}, we make models in
a range of initial masses from 1.7 M$_{\odot}$ to 2.0 M$_{\odot}$ with a
step of 0.02 M$_{\odot}$.  According to the discussion of metallicity
presented in Section~\ref{sec:nonobs}, we allow the initial heavy element
abundances Z$_{i}$ to range from 0.01 to 0.025 with a step of 0.005 as shown in
Table~\ref{tab:input parameter} via
\begin{equation} \label{eq:Zi}
    \log [\text{Z}/\text{X}]
    \simeq
    [\text{Fe}/\text{H}] + \log[\text{Z}/\text{X}]_{\odot},
\end{equation}
where $[\text{Z}/\text{X}]_{\odot} = 0.023$ \citep{grevesse98}.
All input parameters are shown in Table~\ref{tab:input parameter}. Once the
initial heavy element abundance $Z_{i}$ is determined, the dependence of
initial helium mass fraction $Y_{i}$ on $Z_{i}$ can be set using a linear
helium enrichment expression with the primordial helium abundance $Y_{p}=0.24$
and the slope $\Delta Y/\Delta Z= 2$ \citep{demarque04, basu10a} by the following:
\begin{equation} \label{eq:Yip}
    Y_{i}=Y_{p}+\frac{\Delta Y}{\Delta Z}Z_{i}.
\end{equation}
Finally, the initial hydrogen element abundance X$_{i}$ is obtained via
\begin{equation}
    X_{i}=1-Y_{i}-Z_{i}.
\end{equation}

\subsection{Calculating mode frequencies}

We use the GYRE oscillation code \citep{Townsend13} to calculate the mode
frequencies of the models.
Due to surface effects, there is a systematic offset between the observed and model frequencies \citep[e.g.][]{CD84}. In order to obtain the accurate modeling of solar-like oscillations, we should use a systematic shift to correct the model frequencies.
A number of ways have been proposed to correct the frequencies for the
surface issues (usually called the ``surface term'' correction). The most common way to correct for the surface term for stellar models is the method proposed by Kjeldsen et al. (2008). They note that offset between observed and best model frequencies turns out to be closely fitted by a power law. However, there are issues with many of the surface term corrections, even for main sequence stars; and most models perform very badly in the sub-giant and red-giant region. \citet{Schmitt15} found that the two-term model proposed by \citet{Ball14}
works better than other models across a large portion of the HR diagram, and consequently we adopt that for this work.

\section{Results and discussions}
\label{sec:results}
We show the evolutionary track of our models in Figure~\ref{fig:track}. The figure also
shows the two error boxes. $T_{\rm{eff}}=4766 \pm 200 K$ and $\log g= 2.487 \pm 0.4$ have been
used for the large error box while $T_{\rm{eff}}=4974 \pm 161 K$ and $\log g= 2.662 \pm 0.03$ for the smaller error box.
We find that many models fall in the two error boxes.
In order to obtain the best-fit model for KIC 8263801 from among these models,
we calculated the likelihood for each model and determined the
model with the highest likelihood.
The likelihood function is defined as
\begin{equation} \label{eq:lh}
    \mathcal{L}
    =
    \left(\prod_{i=1}^{n}\frac{1}{\sqrt{2\pi}\sigma_{i}}\right)
    \cdot \exp \left\{ -\frac{\chi^{2}}{2} \right\}
\end{equation}
where
\begin{equation} \label{eq:X2}
    \chi^{2}
    =
    \sum_{i=1}^{n}\left(\frac{q_{i}^{\text{obs}}-q_{i}^{\text{model}}}{\sigma_{i}}\right)^{2}.
\end{equation}
with the quantity $q_{i}^{\text{obs}}$ indicating the observed  $T_{\rm{eff}}$, [\text{Fe}/\text{H}], and frequency $\nu_{n,l}$; while the $q_{i}^{\text{model}}$ corresponds to these values of the model.
The quantity $\sigma_{i}$ represents the observational error of  $q_{i}^{\text{obs}}$.
In the study, we use the function as follows:
\begin{equation} \label{eq:lht}
    \mathcal{L}(T_{\rm{eff}},[\text{Fe}/\text{H}],\nu)=\mathcal{L}_{T_{\text{eff}}}\cdot \mathcal{L}_{[\text{Fe}/\text{H}]}\cdot \mathcal{L}_{\nu}
\end{equation}
where
\begin{equation} \label{eq:lt}
    \mathcal{L}_{T_{\rm{eff}}}
    =
    \frac{1}{\sqrt{2\pi}\sigma_{T_{\rm{eff}}}}
    \cdot
    \exp \left\{
        -\frac{(T_{\text{eff,obs}}-T_{\text{eff,model}})^{2}}{2\sigma_{T_{\rm{eff}}}^{2}}
    \right\}
\end{equation}
\begin{equation} \label{eq:Lz}
    \mathcal{L}_{[\text{Fe}/\text{H}]}
    =
    \frac{1}{\sqrt{2\pi}\sigma_{[\text{Fe}/\text{H}]}}
    \cdot
    \exp \left\{
        -\frac{([\text{Fe}/\text{H}]_{\text{obs}}
        -
        [\text{Fe}/\text{H}]_{\text{model}})^{2}}{2\sigma_{[\text{Fe}/\text{H}]}^{2}}
    \right\}
\end{equation}
\begin{equation} \label{eq:Lnu}
    \mathcal{L}_{\nu}
    =
    \prod_{i=1}^{10} \left(
        \frac{1}{\sqrt{2\pi}\sigma_{i}(\nu)}
        \cdot
        \exp \left\{
            -\frac{(\nu^{\text{obs}}_{i}-\nu^{\text{model}}_{i})^{2}}{2\sigma_{i}^{2}(\nu)}
        \right\}
    \right).
\end{equation}
In the above expression, $\nu_i^{\rm model}$ are the surface-term corrected frequencies of the
models.

The values of likelihood function for model of KIC 8263801 as function of mass,
metallicity, age and radius are shown in Figure~\ref{fig:lh}.
We choose the model
of maximizes $\mathcal{L}$ as a candidate for the best-fit model.  The model
parameters are shown in Table~\ref{tab:bestmodel}.
To clearly compare all of the theoretical frequencies of the best-fit model with
observed frequencies, we show the \'{e}chelle diagram of the best-fit model
in Figure~\ref{fig:elle}. In the figure, open symbols are the model frequencies
corrected for the surface term, the filled symbols refer to the observable frequencies.
Circles are used for $l = 0$ modes and squares for $l = 2$ modes.

Finally, we derive the parameters of star as the likelihood weighted mean and
standard deviation from the models and obtain a  mass  of $M/M_{\odot} = 1.793\pm
0.072$, age of $t=1.48\pm 0.21$ Gyr and radius of $R/R_{\odot} = 10.53\pm 0.28$.
Our best-fit model has an age of 1.596 Gyr, the age as determined from the likelihood weighted
average is not too different, and thus the question arises as to whether this star can be a member of
NGC~6866.
There are a number of results about the age of NGC 6866.
For example,
\citet{Loktin94} derived $t = 0.66$ Gyr;
\citet{Gunes12} obtained an age of $t = 0.8 \pm 0.1$ Gyr by 2MASS photometry;
\citet{Kharchenko05} obtained an age of $t=0.5$ Gyr using isochrone-based procedure;
\citet{Bostanci15} derived $t=0.813\pm 0.05$ Gyr with the metallicity of the
cluster being about the solar value;
\citet{Janes14} derived age $t= 0.705 \pm 0.170$ Gyr; and
\cite{Balona13b} estimated the age as $t = 0.65 \pm 0.1$ Gyr with isochrones of solar composition.
While the age estimates vary considerably, it is clear that the consensus is
the age  of NGC 6866 is less than 1 Gyr. Our age estimate of  KIC 8263801 is
significantly higher than 1~Gyr.  Based on these results, we could conclude
that KIC 8263801 is a non-member of the cluster NGC 6866, which is consistent
with \cite{Balona13b}.

We also derived the stellar parameters of KIC 8263801 using the Bayesian tool PARAM (da Silva et al. 2006; Rodrigues et al. 2014, 2017) and the grid models computed with MESA. The mode and its 68 percent credible intervals of the posterior probability density function (PDF) as errors of parameters are mass M/M$_{\odot}$ =  1.8507$_{-0.0249}^{+0.0721}$, age $t$ = $1.5844 _{-0.1555}^{+0.1046}$ Gyr, radius R/R$_{\odot}$ = 10.5724 $_{-0.1185}^{+0.1414}$ and log $g$ = 2.6544 $_{-0.0048}^{+0.003}$ $dex$. These results from PARAM are consistent with our above analysis (M/M$_{\odot}$ = 1.793 $\pm$ 0.072, age $t$ = 1.48 $\pm$ 0.21 Gyr, radius R/R$_{\odot}$ = 10.53 $\pm$ 0.28) within errors, and log $g$ = 2.6544 $_{-0.0048}^{+0.003}$ $dex$ is consistent with observation log $g$ = 2.487 $\pm$ 0.04 $dex$ within uncertainties.

As mentioned above, and as is clear from Figure~\ref{fig:track}, red giants with inert helium cores
and red giants with helium burning cores occupy a common region in the HR diagram.
Errors in temperature and metallicity determinations make it difficult to
unambiguously determine the evolutionary state, and hence age, of a red giant.
Several asteroseismic tools have been developed to distinguish between the
stars \citep{bedding11,kallinger12,Elsworth17}. From the work presented above, we get the the best fit models of KIC 8263801 through modelling. The best fit model of KIC~8263801 is a red giant star which imply that the star in question is on the ascending branch (hence has
an inert core). We use the technique of \cite{kallinger12} to confirm the
result; the data do not have a high enough signal-to-noise ratio to use
the observed $l=1$ period spacing as suggested by \cite{bedding11}, but have the
signal-to-noise is high enough to be able to determine mode frequencies
making it unnecessary to use the method of \citet{Elsworth17}. The fact that
we have good estimates of radial-mode frequencies makes the method of \citet{kallinger12}
ideal.

\cite{kallinger12} found
that the phase function $\epsilon$ determined around $\nu_{\max}$ could be
used to distinguish between RGB and HeB stars.
\citet{White11} find that the phase shift $\epsilon$ changes with the
evolutionary state of a star, and \citet{kallinger12} pointed out especially the
central value of $\epsilon$, which given by the three radial modes around
$\nu_{\max}$, contains the necessary information.
In this work, we tested all of the above methods and find that considering the central radial modes like \citet{kallinger12} is suitable. \citep{kallinger12} expressed the three central radial orders p modes by the following formula:
\begin{equation} \label{eq:nuc0}
    \nu_{c0}=\Delta \nu_{c}(n+\epsilon_{c}^{'}),
\end{equation}
where $\nu_{c0}$ is central frequency varying in the range $\nu_{\max}\pm
0.55\Delta \nu$ and $\Delta \nu_{c}$ is the separation between the three
central modes. As explained by \citet{Rodrigues17}, to determine the
large frequency properly,
we use weighted least squares fit to calculate an average
large-frequency separation $\langle\Delta \nu\rangle$.
We use a Gaussian function, as
described in \cite{mosser12} and \cite{Rodrigues17}, to calculate the
individual weights:
\begin{equation} \label{eq:weight}
    \omega
    =
    \exp \left\{ -\frac{(\nu-\nu_{\max})^{2}}{2\sigma^{2}} \right\}
\end{equation}
 where $\sigma=0.66 \cdot \nu_{\max}^{0.88}$.

If $\nu_{c0}$ and $\Delta \nu_{c}$ are given, the phase shift of
$\epsilon_{c}$ can be determined by the following formula defined by \citet{kallinger12}:
\begin{equation}
    \epsilon_{c}=
    \left\{
    \begin{aligned}
    &\epsilon_{c}^{'}+1 &   & \text{if } \epsilon_{c}^{'} < 0.5 \text{ and }    \Delta\nu > 3 \mu\text{Hz} \\
    &\epsilon_{c}^{'}&     & \text{otherwise}
    \end{aligned}
 \right.
\end{equation}
with
\begin{equation} \label{eq:eps2}
    \epsilon_{c}^{'}
    =
    \left( \frac{\nu_{c0}}{\Delta\nu_{c}} \right) \mod \;  1.
\end{equation}

In Figure 6, we show the central value of $\epsilon_{c}$ for all the models from red giants to red clump (with $\Delta\nu_{c}\leq 8 \mu Hz$) along the evolutionary track, which are calculated using the input parameters in Table 4. Figure 6 display the $\epsilon_{c}$ against the large separation between the central modes.
The stars clearly divide into different groups, where the
circles are RGB models while the triangles are HeB models.  The (blue) square in the
figure denotes the observed value of $\epsilon_{c}$ with error bar of the
star, obtained by error propagation analysis.  We thus find that the KIC 8263801 is indeed an inert-core RGB star.
Although \cite{Abedigamba16} had used the median gravity-mode period spacing $\Delta P$ of $l=1$ modes
to determine that KIC 8263801 is a helium-burning secondary red-clump (SRC) star,
we believe that the low SNR of the data affected the value of $\Delta P$ obtained by them.
Given the mass of the best-fit model, $\Delta P$ would
be the most reliable way of determining the evolutionary state of this
star if measured properly \citep[see][]{bedding11}.
However, KIC 8263801 has a $\Delta\nu_{c}$ of about $5.35 \mu Hz$, and in that
region of Figure~\ref{fig:eps}, the two stages of evolution are well separated.

\section{Conclusions}
\label{sec:conc}

We have done an asteroseismic study of the star KIC~8263801 in order to determine
its age, evolutionary status and membership or otherwise in the open cluster NGC~6866.
Our best fit models has mass $M/M_{\odot} = 1.76$, age $t=1.596$ Gyr,
radius $R/R_{\odot} = 10.3483$  while a full grid analysis gives $M/M_{\odot} = 1.793\pm 0.072$, age $t=1.48\pm 0.21$ Gyr and
radius $R/R_{\odot} = 10.53\pm 0.28$.

The best fit model of KIC~8263801 is on the ascending part of the red giant
branch, making it likely that KIC~8263801 is also in that state. We confirm this
using the $\epsilon_{c}-\Delta\nu_{c}$ diagram.

The age estimates of KIC~8263801 makes it unlikely to be a member cluster NGC~6866,
despite being in the same field since NGC~6866 has age estimates below 1~Gyr. This
result is consistent with that of \citet{Balona13b}.

\acknowledgments

Y.K.T. and N.G.\ acknowledges the research grant from the National Natural
Science Foundation of China (Grant No. 11673005).  S.B.\ acknowledges NSF grant
AST-1514676 and NASA grant NNX16AI09G.  E.P.B.\ acknowledges support from the
European Research Council under the European Community's Seventh Framework
Programme (FP7/2007-2013) / ERC grant agreement no 338251 (StellarAges) and the
National Physical Science Consortium (NPSC) Fellowship.

\begin{table}
\caption{Asteroseismic observational frequencies of KIC 8263801.}
\label{tab:astero-fre}
\begin{center}
\begin{tabular}{c c c c }
\hline\hline
 $n$ &$l$&Value&Error\\
\hline
8 & 0 &   43.242 &   0.068 \\
9& 0 & 48.294 & 0.020  \\
10 & 0 & 53.705 & 0.017 \\
11& 0 & 59.126 & 0.012  \\
12 & 0 & 64.502 & 0.063  \\
\hline
8 & 2 & 41.64 & 0.86 \\
9&2 &47.573 & 0.032  \\
10 & 2 & 52.810 &   0.069\\
11& 2 &   58.352   & 0.020 \\
12& 2 & 63.75 & 0.12 \\
\hline\hline
\end{tabular}
\end{center}
\end{table}

\begin{table}
\begin{center}
\caption{Asteroseismic global parameters of KIC 8263801.}
\label{tab:astero-obs}
\begin{tabular}{c c c }
\hline\hline
 Observable Variable &Value&Source\\
\hline
Large separation $\Delta\nu$  & $5.35 \pm 0.01$  & (1) \\
                              & 5.3541  & (2)  \\
\hline
$\nu_{\max}$&  $56.2 \pm 0.4$ &(1)  \\
&  56.422 & (2) \\
\hline\hline
\end{tabular}

References. (1) this paper; (2)\cite{Abedigamba16}.
\end{center}
\end{table}

\begin{table}
\begin{center}
\caption{Non-asteroseismic data of KIC 8263801.}
\label{tab:classical-obs}
\begin{tabular}{c c c }
\hline\hline
 Observbed Variable &Value&Source\\
\hline
Effective temperature $T_{\rm{eff}} (K)$ &4766  &(1)  \\
                                  &$4974\pm161$  & (2) \\
\hline
Metallicity $[\text{Fe}/\text{H}]$& 0.016 & (1) \\
                      & $0.016\pm0.30$ & (2) \\
\hline
Log$g$ ($dex$)&2.487 & (1) \\
        &$2.662\pm 0.03$ &   (2) \\
\hline\hline
\end{tabular}

References. (1) \citet{brown11} ; (2) \citet{huber14}.
\end{center}
\end{table}

\begin{table*}
\caption{Input parameters for  models}
\label{tab:input parameter}
\begin{center}
\begin{tabular}{c c c c }
\hline\hline
 Variable &Minimum Value&Maximum Value&Step Size\\
\hline
Initial Mass (M$_\odot$)&1.70&2.00&0.02\\
Initial heavy element abundance $Z_i$&0.01&0.025&0.005\\
\hline\hline
\end{tabular}
\end{center}
\end{table*}

\begin{table}
\begin{center}
\caption{The parameters of the best-fit model.}
\label{tab:bestmodel}
\begin{tabular}{c c c }
\hline\hline
 Model parameters &Value  \\
\hline
Mass ($M_{\odot}$)  & 1.76      \\
\hline
Age (Gyr) & 1.596    \\
\hline\hline
Effective temperature $T_{\rm{eff}} (K)$ &4884.3   \\
Luminosity $\log L/L_{\odot}$& 1.7384   \\
$\log g$ (dex) & 2.654   \\
$\log R/R_{\odot}$ & 1.01487 \\
\hline\hline
\end{tabular}
\end{center}
\end{table}

\newpage
\clearpage
\begin{figure}
\epsscale{1.0} \plotone{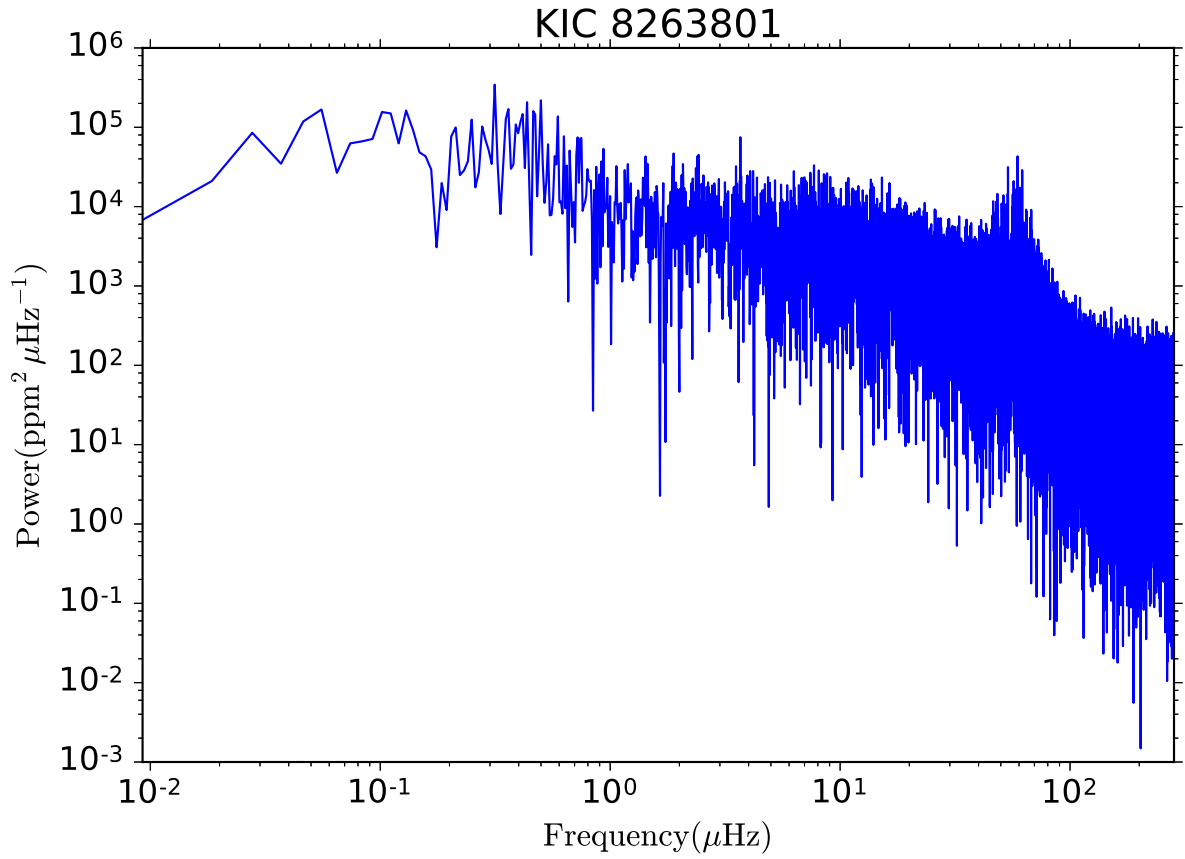}
    \caption{The power spectrum for KIC8263801 from \emph{KEPLER} photometry.}
    \label{fig:psd}
\end{figure}

\newpage
\clearpage
\begin{figure}
\epsscale{1.0} \plotone{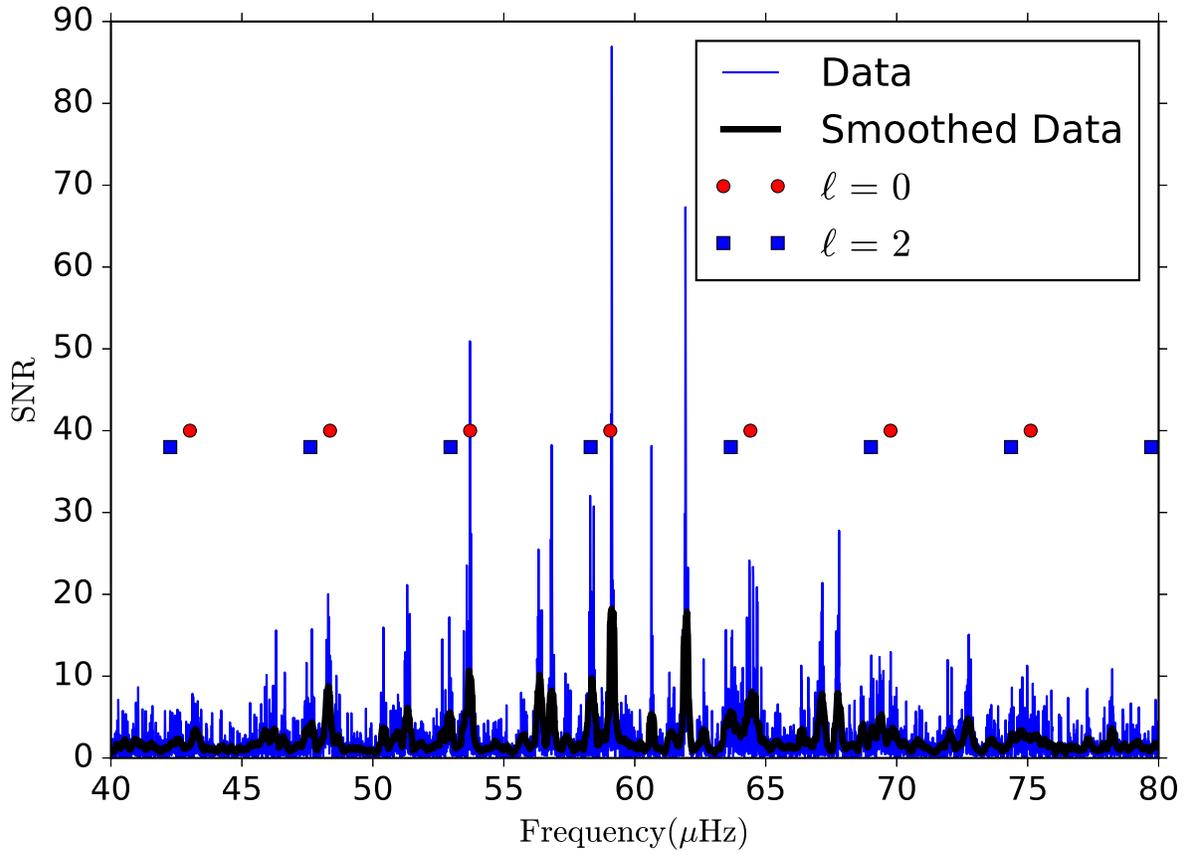}
    \caption{Identified p-mode oscillation spectrum of the red giant KIC8263801 in function of frequency. The (blue) thin
and (black) thick lines denote the power spectrum
before and after smoothing. }
    \label{fig:snr}
\end{figure}

\newpage
\clearpage
\begin{figure}
\epsscale{1.0} \plotone{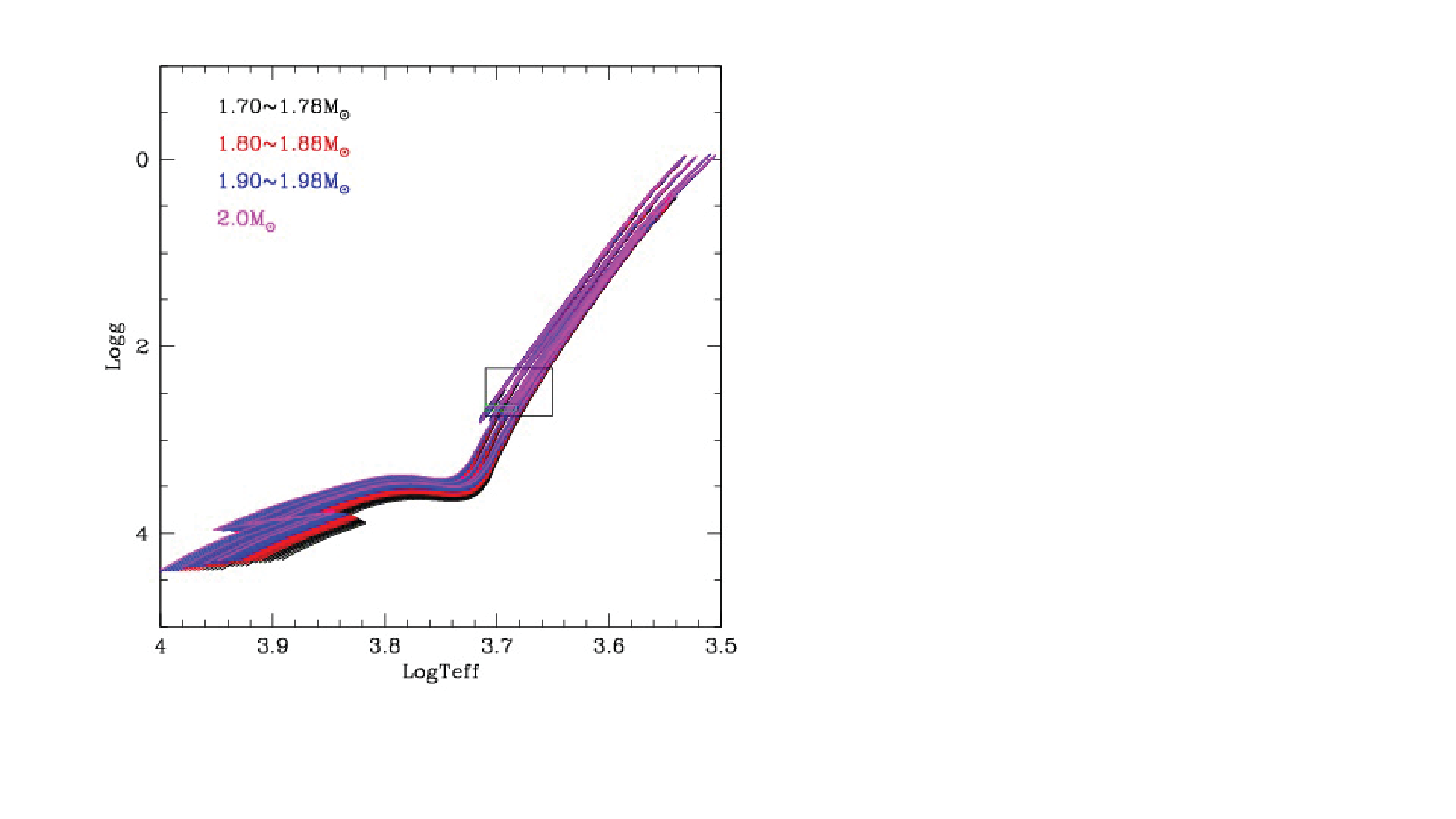}
    \caption{Evolutionary tracks of models constructed for this
work. The large (black) error box corresponds to observational values from the {\it Kepler} website,
while small (green) error box corresponds to the revised observational value given by \cite{huber14} }
    \label{fig:track}
\end{figure}

\newpage
\clearpage
\begin{figure}
\epsscale{1.0} \plotone{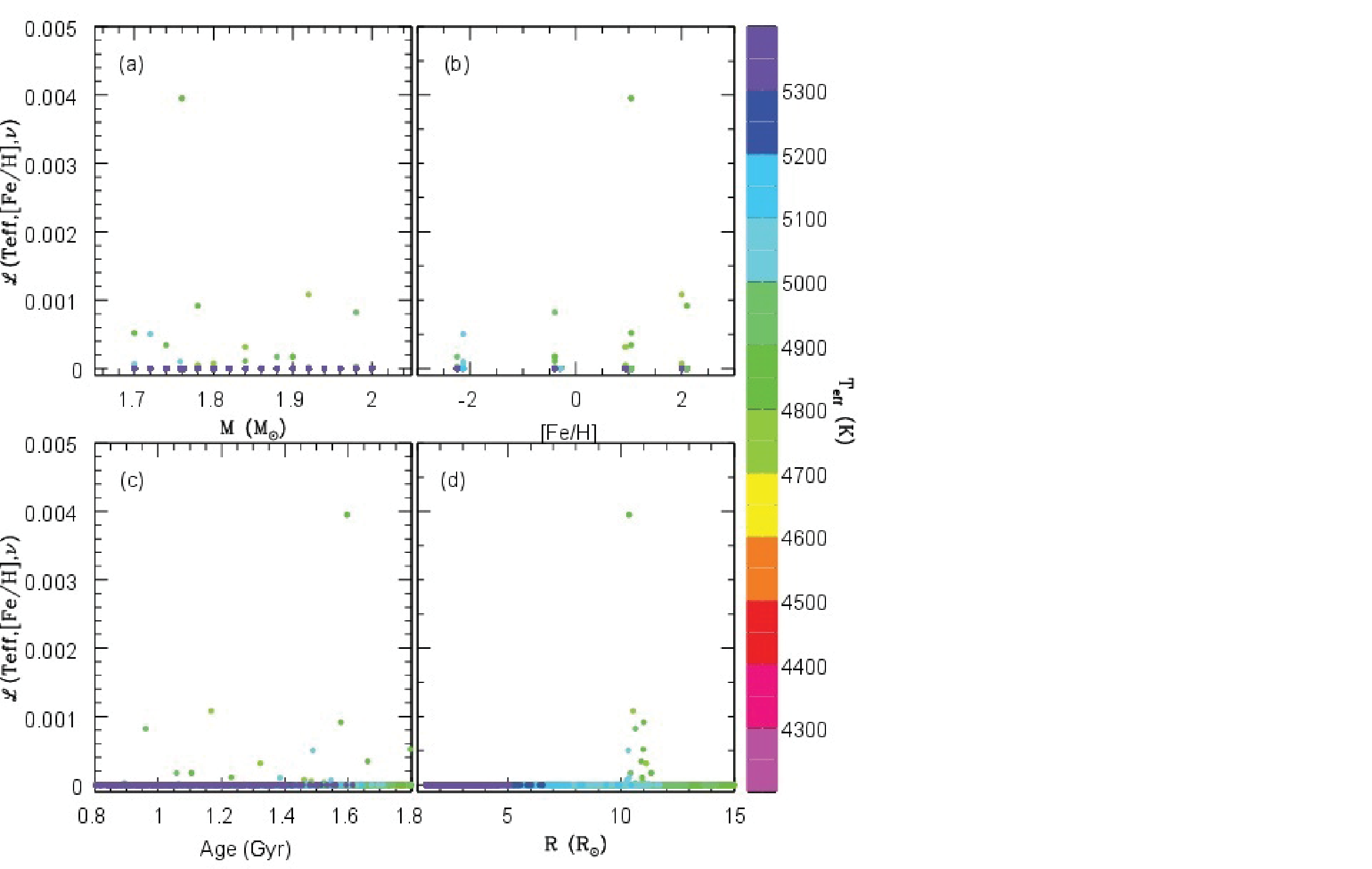}
    \caption{The likelihood values for models of KIC 8263801 as function of mass, metallicity, age and radius, corresponding to panel (a), (b), (c) and (d) respectively.}
    \label{fig:lh}
\end{figure}

\newpage
\clearpage
\begin{figure}
    \epsscale{1.0} \plotone{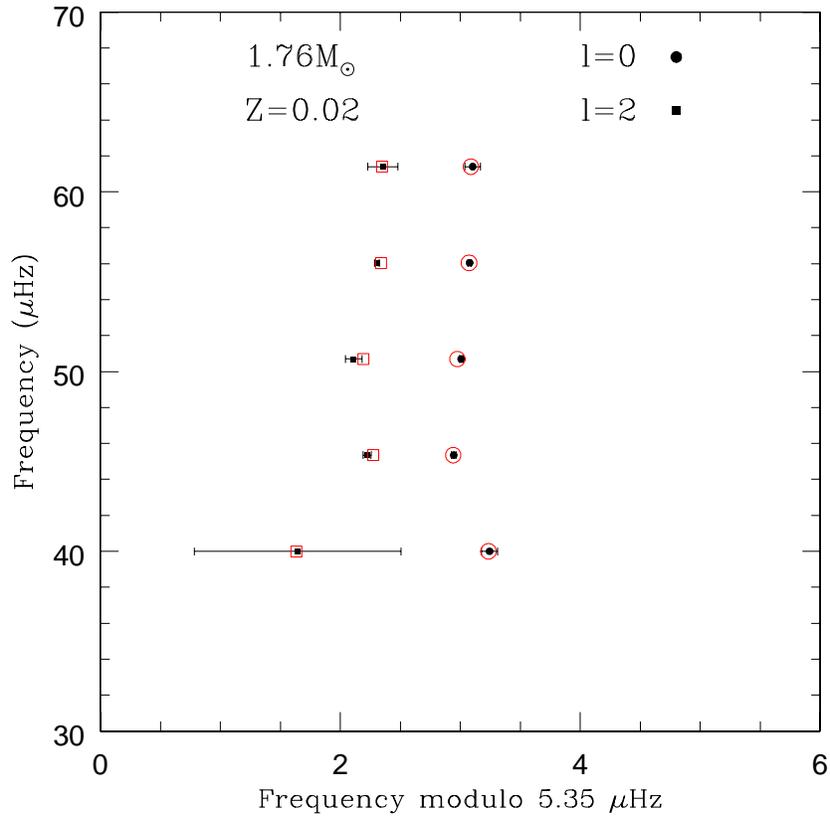}
    \caption{\'{E}chelle diagrams for the best-fit model. Open symbols are the surface-term
corrected frequencies of the best-fit model; filled symbols refer to the observed frequencies.
Circles are used for $l$ = 0 modes, squares for $l$ = 2 modes.}
    \label{fig:elle}
\end{figure}

\newpage
\clearpage
\begin{figure}
    \epsscale{1.0} \plotone{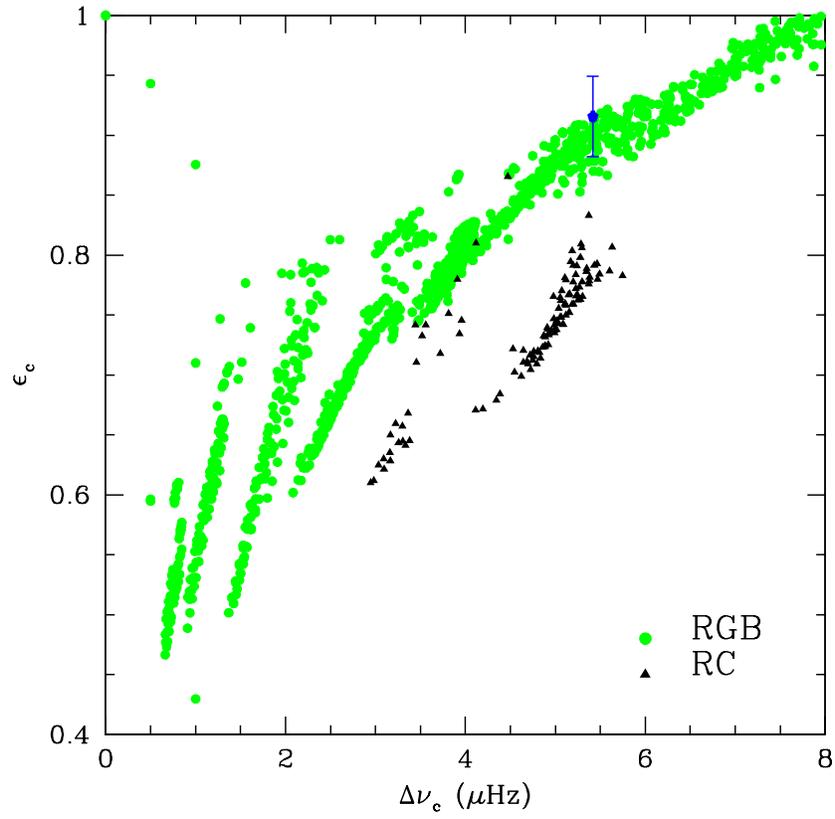}
    \caption{Central value of $\epsilon_{c}$ for models  plotted against the
large separation between the central modes. The (green) filled circles are RGB
models, while the (black) filled triangles are RC models. The (blue) filled
square with error-bars is observed result for KIC 8263801.}
    \label{fig:eps}
\end{figure}


\begin{thebibliography}{}

\bibitem[Abedigamba(2016)]{Abedigamba16}
Abedigamba, O. P., 2016, New Astronomy, 46, 21

\bibitem[Basu et al.(2010a)]{basu10a}
Basu, S., Chaplin, W. J., \&Elsworth, Y. 2010a, ApJ, 710, 1596
%
\bibitem[Basu et al.(2010b)]{basu10b}
Basu, S., Chaplin, W. J., \& Elsworth, Y. 2010b, Ap\&SS, 328, 79

\bibitem[Basu et al.(2011)]{basu11}
Basu, S., Grundahl, F., Stello, D. et al. 2011, ApJ, 729, 10

\bibitem[Ball \& Gizon(2014)]{Ball14}
Ball, W. H. \& Gizon, L. 2014, A\&A, 568, A123

\bibitem[Balona et al.(2013a)]{Balona13a}
Balona, L. A., Medupe, T., Abedigamba, O. P. et al. 2013a, MNRAS, 430, 3472

\bibitem[Balona et al.(2013b)]{Balona13b}
Balona,L.A., Joshi,S., Joshi,Y.C., Sagar,R. 2013b, MNRAS, 429,1466.

\bibitem[Bellinger et al.(2016)]{Bellinger16}
Bellinger, E. P., Angelou, G. C., Hekker, S., et al. 2016, ApJ, 830, 31

\bibitem[Bellinger et al.(2017)]{Bellinger17}
Bellinger, E. P., Angelou, G. C., Hekker, S., et al. 2017, Seis.~Sun~\&~Dist.~Stars, 160, id.05003


\bibitem[Bedding et al.(2010)]{bedding10}
Bedding, T. R., Huber, D., Stello, D. et al. 2010, ApJL, 713, L176
%
\bibitem[Bedding et al.(2011)]{bedding11}
Bedding, T. R., Mosser, B., Huber, D. et al. 2011, Nature, 471, 608

\bibitem[Bedding(2014)]{bedding14}
Bedding, T. R., 2014. Aste.book.60.
%
\bibitem[Beck et al.(2011)]{beck11}
Beck, P. G., Bedding, T. R., Mosser, B., Stello, D. et al. 2011, Science, 332, 205

\bibitem[Borucki et al.(2009)]{Borucki09}
Borucki,W., Koch,D., Batalha,N., 2009, IAUSymposium 289.

\bibitem[Borucki et al.(2010)]{Borucki10}
Borucki, William J.;   Koch, David.; Basri, Gibor. et al., 2010, Science, 327, 977

\bibitem[Bostanci et al.(2015)]{Bostanci15}
Bostanci, Z. F., Ak, T., Yontan, T., et al. 2015, MNRAS, 453, 1095

\bibitem[Brown et al.(1991)]{brown91}
Brown, T. M., Gilliland, R. L., Noyes, R. W., Ramsey, L. W., 1991, ApJ, 368, 599

\bibitem[Brown et al.(2011)]{brown11}
Brown, Timothy M.; Latham, David W.; Everett, Mark E.; Esquerdo, Gilbert A. 2011, AJ, 142, 112

\bibitem[Chaplin et al.(2011)]{Chaplin11}
Chaplin, W. J., Kjeldsen, H., Christensen-Dalsgaard, J., et al. 2011, Sci,
332, 213

\bibitem[Chaplin et al.(2014a)]{Chaplin14}
Chaplin, W. J., Basu, S., Huber, D., et al., 2014a, ApJS, 210, 1

\bibitem[Chaplin et al.(2014b)]{Chaplin14b}
Chaplin, W. J., Elsworth, Y., Davies, G. R., et al. 2014b, MNRAS, 445, 946
%
\bibitem[Christensen-Dalsgaard(1984)]{CD84}
Christensen-Dalsgaard, J. 1984, Space Research in Stellar Activity and Variability, 11

\bibitem[Cox \& Giuli(1968)]{Cox68}
Cox, J. P., \& Giuli, R. T. 1968, Principles of Stellar Structure (New York:
Gordon and Breach

\bibitem[da Silva(2006)]{da06}
da Silva, L., Girardi, L., Pasquini, L. et al. 2006, A\&A, 458, 609

\bibitem[Demarque(2004)]{demarque04}
Demarque, P., Woo, J. -H., Kim, Y. -C. \& Yi, S. K. 2004, ApJS, 155, 667
%
\bibitem[De Ridder et al.(2009)]{de ridder09}
De Ridder, J., Barban, C., Baudin, F. et al. 2009, Nature, 459, 398
%
\bibitem[Davies et al.(2016)]{Davies16a}
Davies, G. R., Aguirre, V. S., Bedding, T. R., et al. 2016, MNRAS, 456, 2183

\bibitem[Davies \& Miglio(2016)]{Davies16b}
Davies, G. R. \& Miglio, A. 2016, Astronomische Nachrichten, 337, 774

\bibitem[Elsworth et al.(2017)]{Elsworth17}
Elsworth Y., Hekker S., Basu S., Davies G., 2017, MNRAS, 466, 3344
%
\bibitem[Gai et al.(2011)]{gai11}
Gai, N., Basu, S., Chaplin, William, J. \& Elsworth, Y. 2011, \apj, 730, 63

\bibitem[Gilliland et al.(2010)]{gilliland10}
Gilliland, R. L., Brown, T. M., Christensen-Dalsgaard, J. et al. 2010, PASP, 122, 131

\bibitem[Grevesse \& Sauval(1998)]{grevesse98}
Grevesse, N. \& Sauval, A. J. 1998, Space Sci. Rev., 85, 161

\bibitem[G\"{u}nes et al.(2012)]{Gunes12}
G\"{u}nes O., Karatas Y., Bonatto C., 2012, NewA, 17, 720

\bibitem[Gualme et al.(2016)]{Gualme16}
Gualme, P., McKeever, J., Jackiewicz, J. et al. 2016, \apj, 832, 121

\bibitem[Guggenberger et al.(2016)]{Guggenberger16}
Guggenberger, E., Hekker, S., Basu, S. et al. 2016, MNRAS, 460, 4277

\bibitem[Guggenberger et al.(2017)]{Guggenberger17}
Guggenberger, E., Hekker, S., Angelou, G. et al. 2016, MNRAS, 470, 2069

\bibitem[Handberg \& Lund(2014)]{Handberg14}
 Handberg, R.\& Lund, M. N. 2014, MNRAS, 445, 2698

 \bibitem[Handberg \& Lund(2017)]{Handberg17}
 Handberg, R.\& Lund, M. N. 2017, A\&A, 597, id.A36

\bibitem[Huber et al.(2009)]{huber09}
Huber, D., Stello, D., Bedding, T. R., et al. 2009, CoAst, 160, 74

\bibitem[Huber et al.(2011)]{huber11}
Huber, D., Bedding, T. R., Stello, D.,  et al. 2011, ApJ, 743, 143

\bibitem[Huber et al.(2014)]{huber14}
Huber, D., Silva Aguirre, V., Matthews, J. M., et al. 2014, ApJS, 211, 2

\bibitem[Hoag et al.(1961)]{Hoag61}
Hoag A. A., Johnson H. L., Iriarte B.,Mitchell R. I., Hallam K. L., Sharpless
S., 1961, Publ. U.S. Naval Obs. (Second Series), 17, 1

\bibitem[Janes et al.(2014)]{Janes14}
Janes K., Barnes S. A., Meibom S., Hoq S., 2014, AJ, 147, 139

\bibitem[Kallinger et al.(2012)]{kallinger12}
Kallinger, T., Hekker, S., Mosser, B. et al. 2012, A\&A, 541, A51
%
\bibitem[Kjeldsen \& Bedding(1995)]{Kjeldsen95}
Kjeldsen, H., Bedding, T. R., 1995, A\&A, 293, 87

\bibitem[Kjeldsen et al.(2008)]{kjeldsen08}
Kjeldsen, H., Bedding, T. R. \& Christensen-Dalsgaard, J. 2008, \apj, 683, L175
%
\bibitem[Koch et al.(2010)]{koch10}
Koch, D. G., Borucki, W. J., Basri, G. et al. 2010, \apj, 713, L79
%
\bibitem[Kharchenko(2005)]{Kharchenko05}
Kharchenko,N.V., Piskunov,A.E., R\"{o}ser,S., 2005, A\&A, 438, 1163.

\bibitem[Loktin et al.(1994)]{Loktin94}
Loktin A. V., Matkin N. V., Gerasimenko T. P., 1994, Astron. Astrophys.
Trans., 4, 153

\bibitem[Mosser et al.(2012)]{mosser12}
Mosser, B., Elsworth, Y., Hekker, S. et al. 2012, A\&A, 537, A30
%
\bibitem[Paxton et al.(2011)]{Paxton11}
Paxton, B., Bildsten, L., Dotter, A., et al. 2011, ApJS, 192

\bibitem[Paxton et al.(2013)]{Paxton13}
Paxton, B., Cantiello, M., Arras, P., et al. 2013, ApJS, 208

\bibitem[Pinsonneault et al.(2014)]{Pinsonneault14}
Pinsonneault, M. H., Elsworth, Y., Epstein, C., et al. 2014, ApJS, 215, 19

\bibitem[Rodrigues et al.(2014)]{Rodrigues14}
Rodrigues T. S., Girardi, L., Miglio, A. et al., 2014, MNRAS, 445, 2758

\bibitem[Rodrigues et al.(2017)]{Rodrigues17}
Rodrigues T. S., Bossini, D., Miglio, A. et al., 2017, MNRAS, 467, 1433

\bibitem[Stello et al.(2011)]{stello11}
Stello, D., Meibom, S., Gilliland, R. L., et al. 2011, ApJ, 739, 13

\bibitem[Schmitt \& Basu(2015)]{Schmitt15}
Schmitt, J. R. \& Basu, S. 2015, ApJ, 808, 123

\bibitem[Silva Aguirre et al.(2015)]{Silva15}
Silva Aguirre, V., Davies, G. R., Basu, S., et al. 2015, MNRAS, 452, 2127

\bibitem[Silva Aguirre et al.(2017)]{Silva17}
Silva Aguirre, V., Lund, M. N., Antia, H. M., et al. 2017, ApJ, 835, 173

\bibitem[Tang et al.(2008a)]{tang08a}
Tang, Y. K., Bi, S.L., Gai, N., \& Xu, H.Y. 2008a, Chinese J. A\&A, 8, 421

\bibitem[Tang et al.(2008b)]{tang08b}
Tang, Y. K., Bi, S. L., \& Gai, N. 2008b, NewA, 13, 541

\bibitem[Tang \& Gai(2011)]{tang11}
Tang, Y. K. \& Gai, N. 2011, A\&A, 526, A35

\bibitem[Tassoul(1980)]{Tassoul80}
Tassoul, M., 1980. ApJS 43, 469.

\bibitem[Townsend \& Teitler(2013)]{Townsend13}
Townsend R. H. D., Teitler S. A., 2013, MNRAS, 435, 3406

\bibitem[White et al.(2011)]{White11}
White T. R., Bedding T. R., Stello D., Christensen-Dalsgaard J., Huber D.,
Kjeldsen H., 2011, ApJ, 743, 161

\bibitem[Wu \& Li(2016)]{Wu16}
Wu, Tao, \& Li, Yan, 2016, ApJ, 818, L13

\bibitem[Yildiz et al.(2017)]{Yildiz17}
 Yildiz, M., \c{C}elik Orhan, Z., \"{o}rtel, S., Roth, M. 2017, MNRAS, 470, L25

\end{thebibliography}
\end{document}